\documentclass[aps,pra,twocolumn,showpacs]{revtex4}
\begin{document}
\title{Optical Generation and Quantitative Characterizations
of Electron-hole Entanglement}
\author{Yu Shi}

\affiliation{Cavendish Laboratory,
University of Cambridge, Cambridge CB3 0HE, United Kingdom}

\affiliation{Department of Physics, University of Illinois at
Urbana-Champaign, Urbana, IL 61801, USA}

Physical Review A {\bf 69}, 032318 (2004)

\begin{abstract}
Using a method of characterizing entanglement in the framework of
quantum field theory, we investigate the optical generation and
quantitative characterizations of quantum entanglement in an
electron-hole system, in presence of spin-orbit coupling, and
especially make a theoretical analysis of a recent experimental
result. Basically, such entanglement should be considered as
between occupation numbers of single particle basis states, and is
essentially generated by coupling between different single
particle basis states in the second quantized  Hamiltonian.
Interaction with two resonant light modes of different circular
polarizations generically leads to a superposition of ground state
and two heavy-hole excitonic states. When and only when the state
is a superposition of only the two excitonic eigenstates, the
entanglement reduces to that between two distinguishable
particles, each with two degrees of freedom, namely, band index,
as characterized by angular momentum, and orbit, as characterized
by position or momentum. The band-index state, obtained by tracing
over the orbital degree of freedom, is found to be a pure state,
hence the band-index and orbital degrees of freedom are separated
in this state. We propose some basic ideas on spatially separating
the electron and the hole, so that the entanglement of
band-indices, or angular momenta, is between spatially separated
electron and hole.
\end{abstract}


\maketitle

\section{Introduction}

As an essential quantum characteristic, quantum entanglement
refers to non-factorization of the state of a composite system in
terms of states of subsystems~\cite{epr}. It is of fundamental
importance for quantum information and quantum
foundations~\cite{book1}. In addition to such systems as photons,
atoms and trapped ions, large amount of work is also going on in
generating entanglement in condensed matter systems.
Investigations are made on generating and separating entangled
electron-electron~\cite{electron} or
electron-hole~\cite{chen,hohenester,bee,bayer} pairs in solid
states.  There is also a lot of proposals of using excitons for
quantum information processing~\cite{sham,exciton,johnson}. Very
recently, coherent optical control of a biexciton in a quantum dot
is also reported~\cite{li}. It should be noted however that
microscopic electronic entanglement is ubiquitous in many-electron
systems and is closely related to the physical properties of
condensed matter~\cite{shi1}. Nevertheless, current researches on
entanglement generation in solid states, largely in mesoscopic
systems, have some special merits or aims such as controllability
and spatial separation.

For an existing electron-hole pair, one can directly study their
entanglement by using the well-known method for distinguishable
particles. However, this approach has limited validity. This is
because electron-hole pairs are excitations, or quasi-particles,
of the many-particle system, and one needs to consider larger
Hilbert space when their creation and annihilation are involved.
For example, the state generated in \cite{chen}, as well as the
state in the proposal in \cite{bee}, have a ground state
component, in which there is no excited electron or hole at all.
Hence an explanation based on the approach of entanglement of
distinguishable particles is not sufficient.

Therefore we need to understand entanglement in the framework of
quantum field theory. Such an approach was made in \cite{shi1},
where it was applied to investigate entanglement in many-particle
physics. For a system of identical particles, entanglement, as the
correlation beyond permutation symmetry, can be defined in terms
of occupation numbers of different single particle basis states or
modes~\cite{shi3,zanardi,van}.  To generate occupation-number
entanglement, it is the coupling between different single particle
basis states, rather than interaction between particles,  that is
essential. We would like to note that photon
entanglement~\cite{photons} is fundamentally also
occupation-number entanglement, and is thus generated by the
mode-mode coupling. The usual description in terms of
distinguishable particles is valid only as a limiting case,
because there is a degree of freedom, e.g. the direction of
movement, effectively distinguishes the photons, and they become
distinguished after separation.

Electron-hole entanglement is basically an occupation-number
entanglement in the many-electron system, and can be simplified to
entanglement between distinguishable quasi-particles when and only
when there is one electron and one hole in each component of the
state. Indeed, it can be seen that mode-mode coupling underlies
the entanglement generation in \cite{chen,hohenester,bee}. In the
situation studied by \cite{bayer}, the mode corresponds to the
position, hence entanglement can be caused merely by the hopping.

In this paper, we make a theoretical account of the physics
underlying the excellent experimental result in \cite{chen},  and
make a detailed analysis on the entanglement and its generation in
this electron-hole system, with the spin-orbit coupling taken into
account. Basically, the state generated by the coupling with the
two laser fields of different circular polarizations is a coherent
superposition of the ground state and two excitonic eigenstates.
Coulomb interaction makes the biexcitonic state off-resonant, but
this is irrelevant to the necessity of interaction of particles in
generating entanglement between distinguishable particles.
Occupation-number entanglement is still generated if the Coulomb
interaction is negligible, and if only one light mode is present.
When the state does not have the ground state component, the
superposition of the two excitonic eigenstates can be described in
terms of two distinguishable particles. Interestingly, the
band-index state, as obtained by tracing out the orbital degrees
of freedom, i.e. the Bloch wavevectors or the positions, is found
to be a pure state in this case. We briefly propose several
methods to spatially separate the electron and the hole, making
band-index entanglement (i.e. entanglement in angular momenta and
in effective masses) nonlocal in positions.

The rest of this article is organized in the following way. In
Sec.~II, we make an introduction to the method of entanglement
characterization in the framework of quantum field theory. In
Sec.~III, as a preliminary, we discuss electron-hole entanglement
in absence of spin-orbit coupling. Then in Sec.~IV, using field
theory, we give the theoretical account of the physical process
underlying the experiment in Ref.~\cite{chen}. The entanglement in
the resulting state is characterized in Sec.~V. Some basic ideas
about spatially separating the electron and hole are described in
Sec.~VI. A summary is made in Sec.~VII.

\section{Entanglement in quantum field theory}

We first make an overview of the method of entanglement
characterization in quantum field theory, in the setting of
condensed matter physics~\cite{shi1}.

In terms of occupation numbers of single particle states {\em for
a chosen single particle basis}, a many-particle state can be
expressed as
\begin{equation}
|\psi\rangle = \sum_{n_1,\cdots, n_{\infty}}
f(n_1,\cdots,n_{\infty})|n_1,\cdots,n_{\infty}\rangle, \label{pn}
\end{equation}
where  $n_i$ is the occupation number of single particle state
$i$, $|n_1,\cdots,n_{\infty}\rangle \equiv {a_1^{\dagger}}^{n_1}
\cdots {a_{\infty}^{\dagger}}^{n_{\infty}}|0\rangle$.

Choosing a different single particle basis means partitioning the
system into a different set of subsystems, based on which the
entanglement is then defined. But  once a single particle basis is
chosen, the entanglement is invariant under any unitary operation
on individual single particle basis states, i.e. when there is no
coupling between different single particle basis states. In other
words, in the present case,  the meaning of ``local operations''
as previously used in quantum information theory is generalized to
operations on the corresponding single particle basis states, as
indexed by the subscript $i$ above. Of course, it is constrained
that some kinds of generalized ``local'' unitary operations do not
exist physically.

Once this generalization of the meaning of subsystems and local
operations is made, the usual method of calculating the amount of
entanglement, as developed in quantum information theory can be
applied. Quantitatively, one considers the Fock-state reduced
density matrix of a set of single particle basis states $1,\cdots,
l$,
\begin{equation}
\begin{array}{c}
\langle n_1',\cdots,n_l'|\rho_l(1\cdots l)|n_1,\cdots,n_l\rangle
\equiv \nonumber \\
\sum_{n_{l+1},\cdots,n_{\infty}} \langle n_1',\cdots,n_l',
n_{l+1},n_{\infty}|\rho|n_1,\cdots,n_l, n_{l+1},n_{\infty}\rangle.
\end{array}
\label{reduce}
\end{equation}
Its von Neumann entropy measures the entanglement of this set of
single particle basis states and the rest of the system. This is a
use of the well-known result for a pure state of a composite
system, the entanglement between a subsystem $A$ and the rest of
the system is quantified as the von Neumann entropy of the reduced
density matrix of $A$, $S_A=-tr_A \rho_A\ln
\rho_A$~\cite{bennett}.

One can also define the entanglement relative to the ground
state$|G\rangle$, by only considering the  excited particles. Then
$n_i$ in (\ref{reduce}) is understood as the number of  the
excited particles, which are absent in the ground state
$|G\rangle$, i.e. $|n_1,\cdots,n_{\infty}\rangle \equiv
{a_1^{\dagger}}^{n_1} \cdots
{a_{\infty}^{\dagger}}^{n_{\infty}}|G\rangle$.

Now we proceed to dynamics. In general, for a system with two
subsystems $A$ and $B$, the Hamiltonian is always of the form
\begin{equation}
H=H_A+H_B+H_{AB}, \label{hab}
\end{equation}
where $H_A$ only acts on $A$, $H_B$ only acts on $B$, while
$H_{AB}$ acts on both $A$ and $B$. If $H_{AB}=0$, then an initial
non-entangled state $|\phi_A\rangle\otimes|\phi_B\rangle$ evolves
to $\exp(-iH_At)|\phi_A\rangle\otimes\exp(-iH_Bt)|\phi_B\rangle$
at any time $t$, which is still non-entangled. Hence the coupling
term $H_{AB}$ is necessary for entanglement generation. For two
distinguishable particles, $A$ and $B$ can directly represent
these two particles.

Here we consider the non-relativistic field theory. The
Hamiltonian is
\begin{equation}
\begin{array}{ll}
{\cal H} & =  \int
d^3r\hat{\psi}^{\dagger}(\mathbf{r})h(\mathbf{r})
\hat{\psi}(\mathbf{r})+ \int
d^3r\hat{\psi}^{\dagger}(\mathbf{r})h'(\mathbf{r})
\hat{\psi}(\mathbf{r})\nonumber
\\ &  + \frac{1}{2}\int d^3r \int d^3r'
\hat{\psi}^{\dagger}(\mathbf{r})\hat{\psi}^{\dagger}(\mathbf{r}')
V(\mathbf{r},\mathbf{r}')\hat{\psi}(\mathbf{r}')\hat{\psi}(\mathbf{r}),
\label{field} \end{array}
\end{equation}
where $h(\mathbf{r})$ is the single particle Hamiltonian including
the kinetic energy, $h'(\mathbf{r})$ is some external potential
which is not included in $h(\mathbf{r})$, for example, the
coupling with electromagnetic field$, and
V(\mathbf{r},\mathbf{r}')$ is the particle-particle interaction.
The reason for separating $h'$ from $h$ will be clear below. The
field operator $\hat{\psi}(\mathbf{r})$ can be expanded in an
arbitrarily chosen single particle basis as
$\hat{\psi}(\mathbf{r})=\sum_i\phi_i(\mathbf{r})a_i$, where $i$ is
the collective index of the single particle state,   which may
include spin if needed, $\phi_i(\mathbf{r})$ is the single
particle wavefunction in position space.

In the form of (\ref{hab}), the Hamiltonian can be written as
\begin{equation}
\begin{array}{lll}
{\cal H} & = & \sum\limits_{ij} \langle i|h|j\rangle a_i^{\dagger}a_j \\
& & + \sum\limits_{ij}\langle i|h'|j\rangle a_i^{\dagger}a_j +
 \frac{1}{2}\sum\limits_{ijlm} \langle
ij|V|lm\rangle a_i^{\dagger}a_j^{\dagger} a_ma_l, \end{array}
\label{sec}
\end{equation}
Thus the index $i$, denoting single particle basis states, defines
distinguishable subsystems.

Now that the entanglement is that between single particle basis
states, its generation needs, in the Hamiltonian ${\cal H}$,
coupling between different single particle basis states.
Therefore, even if $V=0$ and $h'=0$, as far as $\langle
i|h|j\rangle \neq 0$, ${\cal H}$ can still generate
occupation-number entanglement between single particle basis state
$i$ and $j$. Examples of this case include the tunnelling problem
and hopping between Wannier basis states.

However, in many cases, the  single particle state is defined by
the eigenstates of $h$. For example, electrons and holes
corresponding to band structure, i.e. Bloch states. In this single
particle basis, which we call proper single particle basis,
$h\phi_\mu =\epsilon_{\mu}\phi_\mu$,  $\int
d^3r\hat{\psi}^{\dagger}(\mathbf{r})h(\mathbf{r})
\hat{\psi}(\mathbf{r})= \sum_{\mu} \epsilon_{\mu}
a_{\mu}^{\dagger}a_{\mu}$, whose eigenstates are of the form
$\otimes_{\mu}|n_{\mu}\rangle$, where $\mu$ is the collective
index of the proper single particle basis. Therefore in the proper
single particle basis, entanglement can only be caused by $h'$ or
by $V$ if they couple different modes. Note that only $V$ is
particle-particle interaction.

When there are more than one index in the single particle basis,
one of the indices can be used as the tag effectively
distinguishing the particles, and the other indices determine
whether they are entangled in these degrees of freedom.  With this
{\em effective distinguishability}, the state in the configuration
space of  the remaining degrees of freedom can be directly
obtained from the second-quantized state. For example, in
$\frac{1}{\sqrt{2}}(
a_{\mathbf{k}'\uparrow}^{\dagger}a_{\mathbf{k}\downarrow}^{\dagger}
+
a_{\mathbf{k}'\downarrow}^{\dagger}a_{\mathbf{k}\uparrow}^{\dagger})
|0\rangle$, where $\mathbf{k}'$ and $\mathbf{k}$ represent
momenta, one can say that the particle in $|\mathbf{k}'\rangle$
and the particle in $|\mathbf{k}\rangle$ are spin-entangled. One
can also say that the particle in $|\uparrow\rangle$ and the
particle in $|\downarrow\rangle$ are momentum-entangled. With the
momentum as the distinguishing tag, the state can be written as
$\frac{1}{\sqrt{2}}(|\uparrow\rangle_{\mathbf{k}'}|\downarrow\rangle_{\mathbf{k}}+
|\uparrow\rangle_{\mathbf{k}'}|\downarrow\rangle_{\mathbf{k}})$,
with spin entanglement. Alternatively, with the spin as the
distinguishing tag, the state can be written as
$\frac{1}{\sqrt{2}}(|\mathbf{k}'\rangle_{\uparrow}|\mathbf{k}\rangle_{\downarrow}
+|\mathbf{k}\rangle_{\uparrow}|\mathbf{k}'\rangle_{\downarrow})$,
with momentum entanglement.

\section{Electron-hole entanglement in absence of spin-orbit
coupling}

The ground state of an electron gas is $|G\rangle=
\prod_{\mathbf{k}}^{|{\mathbf{k}}| < k_F}
a_{\mathbf{k}\uparrow}^{\dagger}a_{\mathbf{k}\downarrow}^{\dagger}|0\rangle$,
where $k_F$ is the Fermi momentum. This is clearly a non-entangled
state. One can introduce the hole operator
$b^{\dagger}_{{\mathbf{k}}s}=a_{-{\mathbf{k}}-s}$ for $|
{\mathbf{k}}| <k_F$ (we use $s$ and $-s$ to represent the two spin
states).   An excited state is obtained by creating particle-hole
pairs from the ground state. The state
$a_{{\mathbf{k}}s}^{\dagger}b_{{\mathbf{k}}'s'}^{\dagger}|G\rangle$,
with $|{\mathbf{k}}|>k_F>|{\mathbf{k}'}|$, is separable. But there
is maximal entanglement in state $\frac{1}{\sqrt{2}}(
a_{\mathbf{k}\uparrow}^{\dagger}b_{\mathbf{k}'\downarrow}^{\dagger}
+a_{\mathbf{k}\downarrow}^{\dagger}b_{\mathbf{k}'\uparrow}^{\dagger})
|G\rangle$.  This state  can be written as
 $\frac{1}{\sqrt{2}}(
a_{\mathbf{k}\uparrow}^{\dagger}a_{-\mathbf{k}'\downarrow}^{\dagger}
+
a_{\mathbf{k}\downarrow}^{\dagger}a_{-\mathbf{k}'\uparrow}^{\dagger})
\prod_{|{\mathbf{p}}| < k_F}^{{\mathbf{p}} \neq {-\mathbf{k}'}}
a_{\mathbf{p}\uparrow}^{\dagger}a_{\mathbf{p}\downarrow}^{\dagger}|0\rangle$,
from which it can be seen  that with respect to the empty state,
the entanglement is between the excited electron state  and the
one in the same level as the emptied state but with  opposite
spin. But with respect to the ground state, it is simply
electron-hole entanglement. An electron and a hole, by definition,
correspond to different single particle states, and can be
regarded as distinguishable particles.

Consider one electron is excited from a valence band to a
conduction band.  An eigenstate of this excitation, an exciton, in
the absence of spin-orbit coupling, is
$$\sum_{\mathbf{k},\mathbf{k}'}
A_{\mathbf{k},\mathbf{k}'}|S,S_z\rangle_{\mathbf{k},\mathbf{k}'},$$
where  $|S,S_z\rangle_{\mathbf{k},\mathbf{k}'}$ represents three
triplet states as the ground states,
$|1,1\rangle_{\mathbf{k},\mathbf{k}'} =
a_{\mathbf{k}\uparrow}^{\dagger}b_{\mathbf{k}'\uparrow}^{\dagger}|G\rangle$,
$|1,0\rangle_{\mathbf{k}\mathbf{k}'} =
\frac{1}{\sqrt{2}}(a_{\mathbf{k}\uparrow}^{\dagger}b_{\mathbf{k}'\downarrow}^{\dagger}
-a_{\mathbf{k}\downarrow}^{\dagger}b_{\mathbf{k}'\uparrow}^{\dagger})|G\rangle$
and $|1,-1\rangle_{\mathbf{k},\mathbf{k}'} =
a_{\mathbf{k}\downarrow}^{\dagger}
b_{\mathbf{k}'\downarrow}^{\dagger}|G\rangle$, and one singlet
state $|0,0\rangle_{\mathbf{k},\mathbf{k}'} =
\frac{1}{\sqrt{2}}(a_{\mathbf{k}\uparrow}^{\dagger}b_{\mathbf{k}'\downarrow}^{\dagger}
+a_{\mathbf{k}\downarrow}^{\dagger}
b_{\mathbf{k}'\uparrow}^{\dagger})|G\rangle$.
$A_{\mathbf{k},\mathbf{k}'}$ is determined by the Schr\"{o}dinger
equation in momentum representation,
$(E_{0c}+\hbar^2\mathbf{k}^2/2m_e+E_{0v}+\hbar^2\mathbf{k}^2/2m_h-E)
A_{\mathbf{k},\mathbf{k}'}-\sum_{\mathbf{q},\mathbf{q}'}
(V^{cvvc}_{\mathbf{k}-\mathbf{k}'-\mathbf{q}'\mathbf{q}}-
V^{cvcv}_{\mathbf{k}-\mathbf{q}'\mathbf{q}
-\mathbf{k}'})A_{\mathbf{q},\mathbf{q}'}=0$, where $E_{0c}$ is the
bottom of the conduction electron  band,  $E_{0v}$ is the top of
the valence hole band,
$V^{\mu\nu\sigma\delta}_{\mathbf{k}_1\mathbf{k}_2\mathbf{k}_3\mathbf{k}_4}
=\int
\phi_{\mu\mathbf{k}_1}(\mathbf{r})\phi_{\nu\mathbf{k}_2}(\mathbf{r}')
V(\mathbf{r}-\mathbf{r}')\phi_{\sigma\mathbf{k}_3}(\mathbf{r}')
\phi_{\delta\mathbf{k}_4}(\mathbf{r}) d^3\mathbf{r}
d^3\mathbf{r}'$, $\mu$, $\nu$, $\sigma$ and $\delta$ represent
band indices.

Consider $\sum_{\mathbf{k},\mathbf{k}'}
A_{\mathbf{k},\mathbf{k}'}\frac{1}{\sqrt{2}}
(a_{\mathbf{k}\uparrow}^{\dagger}b_{\mathbf{k}'\downarrow}^{\dagger}
\pm
a_{\mathbf{k}\downarrow}^{\dagger}b_{\mathbf{k}'\uparrow}^{\dagger})|G\rangle$.
The occupation-number entanglement between the electron basis
state $|\mathbf{k},\uparrow(\downarrow)\rangle_e$ and the rest of
the system is $-(\alpha_\mathbf{k}/2)\ln(\alpha_{\mathbf{k}}/2)-
(1-\alpha_\mathbf{k}/2)\ln(1-\alpha_{\mathbf{k}}/2)$,  where
$\alpha_{\mathbf{k}'} =
\sum_{\mathbf{k}}|A_{\mathbf{k},\mathbf{k}'}|^2$. The
occupation-number entanglement between the hole basis state
$|\mathbf{k}',\downarrow(\uparrow)\rangle_h$ and the rest of the
system is $-(\alpha_{\mathbf{k}'}/2)\ln(\alpha_{\mathbf{k}'}/2)-
(1-\alpha_{\mathbf{k}'}/2)\ln(1-\alpha_{\mathbf{k}'}/2)$.

On the other hand, because the electron and the hole are
effectively distinguishable, these states can be written, in the
configuration space, as
$$\sum_{\mathbf{k},\mathbf{k}'}
A_{\mathbf{k},\mathbf{k}'}|\mathbf{k}\rangle|\mathbf{k}'\rangle|S,S_z\rangle.$$
So the orbital and spin degrees of freedom are separated, as
consistent with the presumption that spin-orbit coupling is
neglected. The entanglement in the spin state $|S,S_z\rangle$ is
well-known.

But note that when a state is a superposition of ground state,
where occupation-numbers of the relevant electron and hole states
are zero, and excitonic states, the  entanglement cannot be
treated as entanglement between distinguishable particles.

An electron-hole pair can be generated by electron-light coupling,
which corresponds to $h'$ in (\ref{sec}).  This underlies the
experimental result in \cite{chen}. A theoretical study is made
below, with  spin-orbit coupling taken into account.

\section{Physical Process with Spin-orbit Coupling}

Now we study the physical process underlying the experiment in
\cite{chen}.  We shall consider coupling with lights which are
only relevant to the two conduction bands  $c1$ and $c2$,  and the
two heavy-hole bands $h1$ and $h2$. For these four bands, the
total angular momentum $(j,m_j)$ is, respectively,
$(\frac{1}{2},\frac{1}{2})$, $(\frac{1}{2},-\frac{1}{2})$,
$(\frac{3}{2},\frac{3}{2})$, $(\frac{3}{2},-\frac{3}{2})$. The
neglect of other bands only affects the microscopic expressions of
some effective parameters, and the ground state energy, which is
not relevant. This band structure is a consequence of spin-orbit
coupling, i.e. the spin-orbit coupling has been included in the
one-particle Hamiltonian $h(\mathbf{r})$ as in Eq.~(\ref{field}).
The field operator $\hat{\psi}(\mathbf{r})$ can be expanded as
$\sum_{i\mathbf{k}}
[a_{i\mathbf{k}}\phi_{ci\mathbf{k}}(\mathbf{r})
+b_{i-\mathbf{k}}^{\dagger}\phi_{hi\mathbf{k}}(\mathbf{r})]$,
where $i=1,2$, $a_{i\mathbf{k}}$ is the electron annihilation
operator for the band $ci$, $b_{i-\mathbf{k}}\equiv
a^{\dagger}_{i\mathbf{k}}$ is the hole annihilation operator for
band $hi$. Consequently, the field theoretical Hamiltonian ${\cal
H}_e$ is reduced to
$$\begin{array}{c} {\cal H}_e = E_0 +
\sum\limits_{i}\sum_{\mathbf{k}}E_{ci\mathbf{k}}
a_{i\mathbf{k}}^{\dagger}a_{i\mathbf{k}} +
\sum\limits_i\sum\limits_{\mathbf{k}}E_{hi\mathbf{k}}
b_{i\mathbf{k}}^{\dagger}b_{i\mathbf{k}} \\
+\frac{1}{2}\sum\limits_{ij}\sum\limits_{\mathbf{k}_{\mu}
 \mathbf{k}_{\nu}\mathbf{k}_{\sigma}\mathbf{k}_{\delta}}
V^{cicjcjci}_{\mathbf{k}_{\mu}
 \mathbf{k}_{\nu}\mathbf{k}_{\sigma}\mathbf{k}_{\delta}}
a^{\dagger}_{i\mathbf{k}_{\mu}} a^{\dagger}_{j\mathbf{k}_{\nu}}
a_{j\mathbf{k}_{\sigma}}a_{i\mathbf{k}_{\delta}}\\
+ \frac{1}{2}\sum\limits_{ij}\sum\limits_{\mathbf{k}_{\mu}
 \mathbf{k}_{\nu}\mathbf{k}_{\sigma}\mathbf{k}_{\delta}}
V^{hihjhjhi}_{-\mathbf{k}_{\mu}
-\mathbf{k}_{\nu}-\mathbf{k}_{\sigma}-\mathbf{k}_{\delta}}
b^{\dagger}_{i\mathbf{k}_{\delta}}
b^{\dagger}_{j\mathbf{k}_{\sigma}}
b_{j\mathbf{k}_{\nu}}b_{i\mathbf{k}_{\mu}}-\\
\sum\limits_{ij}\sum\limits_{\mathbf{k}_{\mu}
 \mathbf{k}_{\nu}\mathbf{k}_{\sigma}\mathbf{k}_{\delta}}
(V^{cihjhjci}_{\mathbf{k}_{\mu}
-\mathbf{k}_{\nu}-\mathbf{k}_{\sigma}\mathbf{k}_{\delta}}
 -V^{cihjcihj}_{\mathbf{k}_{\mu}
 -\mathbf{k}_{\nu}\mathbf{k}_{\delta}-\mathbf{k}_{\sigma}})
a^{\dagger}_{i\mathbf{k}_{\mu}}a_{i\mathbf{k}_{\delta}}
b^{\dagger}_{j\mathbf{k}_{\sigma}} b_{j\mathbf{k}_{\nu}},
\end{array}
$$
where $i=1,2$, $j=1,2$, $E_0$ is the ground state energy,
 $V^{\mu\nu\sigma\delta}_{\mathbf{k}_{\mu}
 \mathbf{k}_{\nu}\mathbf{k}_{\sigma}\mathbf{k}_{\delta}}
 =\int
d^3rd^3r'\phi^*_{\mu\mathbf{k}_{\mu}}(\mathbf{r})
\phi^*_{\nu\mathbf{k}_{\nu}}(\mathbf{r}')
V(\mathbf{r}-\mathbf{r}')\phi_{\sigma\mathbf{k}_{\sigma}}(\mathbf{r}')
\phi_{\delta\mathbf{k}_{\delta}}(\mathbf{r})$ is the matrix
element of Coulomb interaction $V$ between single particle states
indexed by band index and Bloch wavevector. Originally there is
degeneracy between the two conduction bands and between the two
heavy-hole bands. But in accordance with the
experiment~\cite{chen}, here it is assumed that the degeneracy has
been removed by coupling with a perturbative magnetic field, which
is included in the single particle Hamiltonian, which also
includes the kinetic energy and spin-orbit coupling. The lifting
of degeneracy allows  the use of effective mass theory for
non-degenerate bands, which says
$E_{ci\mathbf{k}}=E_{ci0}+\hbar^2\mathbf{k}^2/2m_{ci}$ and
$E_{hi\mathbf{k}}=E_{hi0}+\hbar^2\mathbf{k}^2/2m_{hi}$, where
$m_{ci}$ and $m_{hi}$ are effective masses.

Taking into consideration the angular momentum selection rule in
their generation,  the relevant low-lying eigenstates of ${\cal
H}_{e}$ are the following: First, the ground state $|G\rangle$;
second, the single-exciton state made up of an electron in $c1$
band and a hole in $h1$ band,
$$\begin{array}{lll}
|S1\rangle & = & \sum_{\mathbf{k},\mathbf{k}'}
\Phi_1(\mathbf{k},\mathbf{k}')a_{1\mathbf{k}}^{\dagger}
b_{1\mathbf{k}'}^{\dagger}|G\rangle \\
&=&\int\int d\mathbf{r}d\mathbf{r}'
\Psi_1(\mathbf{r},\mathbf{r}')a_{1\mathbf{r}}^{\dagger}
b_{1\mathbf{r}'}^{\dagger}|G\rangle, \end{array}$$
with energy $E_0+E_1$; third, the single-exciton state made up of
an electron in $c2$ band and a hole in $h2$ band,
$$\begin{array}{lll}
|S2\rangle & =& \sum_{\mathbf{k},\mathbf{k}'}
\Phi_2(\mathbf{k},\mathbf{k}')a_{2\mathbf{k}}^{\dagger}
b_{2\mathbf{k}'}^{\dagger}|G\rangle\\ & = & \int\int
d\mathbf{r}d\mathbf{r}'
\Psi_2(\mathbf{r},\mathbf{r}')a_{2\mathbf{r}}^{\dagger}
b_{2\mathbf{r}'}^{\dagger}|G\rangle, \end{array}$$
with energy $E_0+E_2$; fourth, the biexciton state made up of an
electron in $c1$ band, an electron in $c2$ band,  a hole in $h1$
band, and a hole in $h2$ band,
$$
\begin{array}{lll}
|B\rangle &= &
\sum\limits_{\mathbf{k}_1,\mathbf{k}_1',\mathbf{k}_2,\mathbf{k}_2'}
\Phi_{B}(\mathbf{k}_1,\mathbf{k}_1',\mathbf{k}_2,\mathbf{k}_2')
a_{1\mathbf{k}_1}^{\dagger} b_{1\mathbf{k}_1'}^{\dagger}
a_{2\mathbf{k}_2}^{\dagger} b_{2\mathbf{k}_2'}^{\dagger}|G\rangle
\\
&=& \int\int\int\int d\mathbf{r}_1d\mathbf{r}_1'
d\mathbf{r}_2d\mathbf{r}_2'
\Psi_{B}(\mathbf{r}_1,\mathbf{r}'_1,\mathbf{r}_2,\mathbf{r}'_2)\\
& & \times  a_{1\mathbf{r}_1}^{\dagger}
b_{1\mathbf{r}'_1}^{\dagger}a_{2\mathbf{r}_2}^{\dagger}
b_{2\mathbf{r}_2'}^{\dagger}|G\rangle, \end{array}$$
with energy $E_0+E_{B}$. The wavefunctions of these exciton and
biexciton states are the lowest bound state wavefunctions of the
corresponding stationary Schr\"{o}dinger equations.

Now consider the coupling with light. For a light
$\mathbf{A}(\mathbf{r}) = \mathbf{\eta}_{\mathbf{q}}
(A_{\mathbf{q}}
e^{i\mathbf{q}\cdot\mathbf{r}-i\omega_{\mathbf{q}}t} +
A_{\mathbf{q}}^*
e^{-i\mathbf{q}\cdot\mathbf{r}+i\omega_{\mathbf{q}}t})$, where
$\mathbf{\eta}_{\mathbf{q}}$ is the unit polarization vector,
$\mathbf{q}\approx 0$, the electron-light coupling Hamiltonian is
$$\begin{array}{lll}
{\cal H}_{eq} &  = & \int \psi^{\dagger}(\mathbf{r})
d\mathbf{r}[-\frac{e}{m}\mathbf{p}\cdot \mathbf{A}(\mathbf{r})]
\psi(\mathbf{r}) d\mathbf{r} \\
& = & -\frac{e}{m}
\sum_{\mu\nu\mathbf{k}}(M_{\mu\nu\mathbf{k}\mathbf{q}}
a^{\dagger}_{\mu\mathbf{k}}a_{\nu \mathbf{k}}A_{\mathbf{q}}e^{-i
\omega_{\mathbf{q}}t}+H.c.),\end{array} $$
where  $M_{\mu\nu\mathbf{k}\mathbf{q}}=  \int_{cell} d\mathbf{r}
u_{\mu\mathbf{k}}^*(\mathbf{r})
\mathbf{p}\cdot\mathbf{\eta}_{\mathbf{q}}u_{\nu\mathbf{k}}
(\mathbf{r})$. This remains unchanged in presence of a magnetic
field, which is about constant in the crystal unit cell, since
$\int_{cell}
d\mathbf{r}u_{\mu\mathbf{k}}^*(\mathbf{r})u_{\nu\mathbf{k}}=0$.

In the present situation, consider the coupling with two
circularly polarized monochromatic lights~\cite{chen}. One is with
$\mathbf{\eta}_{\mathbf{q}1}=(-\mathbf{e}_x-i\mathbf{e}_y)/\sqrt{2}$,
$M_{\mu\nu\mathbf{k}\mathbf{q}1} \neq 0$ only for $\mu=c1$,
$\nu=h1$. Its interaction with electrons is
$${\cal H}_{eq1} = -\frac{e}{m}
\sum_{\mathbf{k}}(M_{c1h1\mathbf{k}\mathbf{q}1}
a^{\dagger}_{1\mathbf{k}}b_{1-\mathbf{k}}^{\dagger}A_{\mathbf{q}1}e^{-i
\omega_{\mathbf{q}1}t}+H.c.).$$
The other light is  with $\mathbf{\eta}_{\mathbf{q}2}=
(\mathbf{e}_x-i\mathbf{e}_y)/\sqrt{2}$
$M_{\mu\nu\mathbf{k}\mathbf{q}2} \neq 0$ only for $\mu=c2$,
$\nu=h2$. Its interaction with electrons is
$${\cal H}_{eq2} = -\frac{e}{m}
\sum_{\mathbf{k}}(M_{c2h2\mathbf{k}\mathbf{q}2}
a^{\dagger}_{2\mathbf{k}}b_{2-\mathbf{k}}^{\dagger}A_{\mathbf{q}2}e^{-i
\omega_{\mathbf{q}2}t}+H.c.).$$

With the interaction with these two light modes, the total
Hamiltonian is
\begin{equation}
{\cal H}= {\cal H}_e+{\cal H}_{eq1}+{\cal H}_{eq2}.
\end{equation}
Under ${\cal H}$, the electronic state $|\Psi(t)\rangle$ is
determined by
\begin{equation}
i\hbar \frac{\partial |\Psi(t)\rangle}{\partial t} = {\cal H}
|\Psi(t)\rangle. \label{sc}
\end{equation}
In terms of the four relevant eigenstates of ${\cal H}_e$,
$|\Psi(t)\rangle$ can be expanded as
$$\begin{array}{ll} |\Psi(t)\rangle
 = &  f_G |G\rangle+ f_{S1}e^{-iE_1 t/\hbar}|S1\rangle
  + f_{S2}e^{-iE_2 t/\hbar}|S2\rangle
  \\& + f_{B}e^{-iE_{B} t/\hbar}|B\rangle, \end{array} $$ where
the coefficients $f_k$ are determined by
\begin{equation}
i\hbar\frac{\partial f_k(t)}{\partial t} = \sum_n f_n(t)
e^{-i(E_n-E_k)t/\hbar}(\langle k|{\cal H}_{eq1}|n\rangle+\langle
k| {\cal H}_{eq2}|n\rangle),
\end{equation}
as  obtained from (\ref{sc}). The initial condition is $f_G(0)=1$.

Note that the only non-vanishing matrix elements of $H_{eq11}$ and
$H_{eq2}$ are
$$\langle S1|{\cal H}_{eq1}|G\rangle=
-\frac{e}{m}\sum_{\mathbf{k}}
\Phi^*_1(\mathbf{k},-\mathbf{k})M_{c1h1\mathbf{k}
\mathbf{q}1}A_{\mathbf{q}1}e^{-i\omega_{\mathbf{q}1}t},$$
$$\langle S2|{\cal H}_{eq2}|G\rangle=
-\frac{e}{m}\sum_{\mathbf{k}}
\Phi^*_2(\mathbf{k},-\mathbf{k})M_{c2h2\mathbf{k}
\mathbf{q}2}A_{\mathbf{q}2}e^{-i\omega_{\mathbf{q}2}t},$$
$$\begin{array}{ll}
\langle B|{\cal H}_{eq2}|S1\rangle & =
-\frac{e}{m}\sum_{\mathbf{k},\mathbf{k}_1,\mathbf{k}_1'}
\Phi^*_B(\mathbf{k}_1,\mathbf{k}_1',\mathbf{k},-\mathbf{k})\\
& \times\Phi_1(\mathbf{k}_1,\mathbf{k}_1')
M_{c2h2\mathbf{k}\mathbf{q}2}A_{\mathbf{q}2}e^{-i\omega_{\mathbf{q}2}t},
\end{array}$$
$$\begin{array}{ll}
\langle B|{\cal H}_{eq1}|S2\rangle=&
-\frac{e}{m}\sum_{\mathbf{k},\mathbf{k}_2,\mathbf{k}_2'}
\Phi^*_B(\mathbf{k},-\mathbf{k},\mathbf{k}_2,\mathbf{k}_2')\\
&\times\Phi_2(\mathbf{k}_2,\mathbf{k}_2')
M_{c1h1\mathbf{k}\mathbf{q}1}A_{\mathbf{q}1}e^{-i\omega_{\mathbf{q}1}t},
\end{array}$$
and their transposes.

In the perturbative expansion, $f_k=\sum_j f_k^{(j)}$, where $j$
represents the order of perturbation. Thus, $i\hbar\partial
f_k^{(j)}(t)/\partial t =\sum_n \exp[-i(E_n-E_k)t/\hbar](\langle
k|{\cal H}_{eq1}|n\rangle+\langle k| {\cal
H}_{eq2}|n\rangle)f_n^{(j-1)}$. Therefore, in each term of
$f^{(j)}$, there should be a product of $j$ matrix elements of
${\cal H}_{eq1}$ or ${\cal H}_{eq2}$, in terms of $j-1$
intermediate states connecting initial state $|G\rangle$ and state
$|k\rangle$., i.e. $\langle k|{H}_{1}|n_1\rangle \langle
n_1|{H}_2|n_2\rangle\cdots \langle n_{j-1}|{H}_{k}|G\rangle$,
where each $H_i$ ($i=1,\cdots k$) is either  ${\cal H}_{eq1}$ or
 ${\cal H}_{eq2}$.

From this, it can be seen that $f_B(t)$ approaches zero quickly
with time. First, due to angular momentum selection rule, $f_B(t)$
exactly vanishes in odd orders, where there must be $\langle
B|{\cal H}_{eq1}+{\cal H}_{eq2}|G\rangle$, which is zero. Second,
it can be seen that given $\hbar\omega_{\mathbf{q}i}=E_i$ while
$E_B \neq E_1+E_2$ due to Coulomb interaction, each even order,
involving integrals over time, approaches zero quickly with time.
This is, of course, the off-resonance effect. In contrast,
$f_{Si}(t)$ is nonvanishing and becomes appreciable for
sufficiently long time, because of resonance
$\hbar\omega_{\mathbf{q}i}=E_i$. The first order result is
$f_{S1}(t) \approx \frac{e}{\hbar
m}\sum_{\mathbf{k}}\Phi^*_1(\mathbf{k},-\mathbf{k})M_{c1h1\mathbf{k}
\mathbf{q}1}A_{\mathbf{q}1}\frac{e^{i(E_1/\hbar-\omega_{\mathbf{q}1})t}-1}
{\omega_{G,S1}-\omega_{\mathbf{q}1}}$ (and the similar expression
for $f_{S2}$), which can easily lead to the well-known Eliot
formula of the transition rate, which is usually derived in a
different way.

The point we particularly want to emphasize is that although
Coulomb interaction prevents the appearance of $|B\rangle$,  it is
irrelevant to the situation that interaction of particles is
needed to generate entanglement between distinguishable particles.
The entanglement in the present case is that of occupation
numbers, whose generation depends on the coupling between single
particle basis states, which is offered here by the electron-light
interaction. In fact, occupation-number entanglement still exists
even when the state is a superposition of the ground state and one
single-excitonic eigenstate, or even simply in a single-excitonic
eigenstate. If Coulomb interaction is negligible, the states
generated by the two lights are two independent states, each being
a superposition state of the ground state and an single-excitonic
eigenstate, in which there exists occupation-number entanglement.

In general, interactions of particles is not necessary unless the
single particle basis states are the eigenstates of the single
particle Hamiltonian. In the case of optical control, the single
particle Hamiltonian which defines the single particle basis does
not include the electron-light interaction, which thus couples
different single particle basis states.

\section{Entanglement characterization of the superposition of
different excitonic states}

In the preceding section, we have shown that the state generated
must be of the form
\begin{equation}
|\Psi\rangle  =   g_G |G\rangle+ g_{S1}|S1\rangle +
g_{S2}|S2\rangle, \label{ptt}
\end{equation}
which was obtained in  the experiment, as indicated by the
interference line shape in the coherent nonlinear response, using
the two light modes as pump and probe fields
respectively~\cite{chen}.

Now we analyze the entanglement in the state $|\Psi\rangle$ in
Eq.~(\ref{ptt}). In details of the occupation-numbers of the
single electron or hole basis states at the four relevant bands,
$$\begin{array}{lll}
|\Psi\rangle & = & g_G\prod\limits_{\mathbf{k}_1\mathbf{k}'_1}
|0\rangle_{c1\mathbf{k}_1}|0\rangle_{h1\mathbf{k}'_1}
\prod\limits_{\mathbf{k}_2\mathbf{k}'_2}
|0\rangle_{c2\mathbf{k}_2}|0\rangle_{h2\mathbf{k}'_2} \\
&&+g_{S1}\sum\limits_{\mathbf{k}_1\mathbf{k}'_1}\Phi_1(\mathbf{k}_1,\mathbf{k}'_1)
|1\rangle_{c1\mathbf{k}_1}|1\rangle_{h1\mathbf{k}'_1}
\\
&&\times
\prod\limits_{\overline{\mathbf{k}}_1\overline{\mathbf{k}}'_1}
|0\rangle_{c1\overline{\mathbf{k}}_1}|0\rangle_{h1\overline{\mathbf{k}}'_1}
\prod\limits_{\mathbf{k}_2\mathbf{k}'_2}
|0\rangle_{c2\mathbf{k}_2}|0\rangle_{h2\mathbf{k}'_2}\\
&&+g_{S2}\prod\limits_{\mathbf{k}_1\mathbf{k}'_1}
|0\rangle_{c1\mathbf{k}_1}|0\rangle_{h1\mathbf{k}'_1}\\
&&\times
\sum\limits_{\mathbf{k}_2\mathbf{k}'_2}\Phi_2(\mathbf{k}_2,\mathbf{k}'_2)
|1\rangle_{c2\mathbf{k}_2}|1\rangle_{h2\mathbf{k}'_2}
\prod\limits_{\overline{\mathbf{k}}_2\overline{\mathbf{k}}'_2}
|0\rangle_{c2\overline{\mathbf{k}}_2}|0\rangle_{h2\overline{\mathbf{k}}'_2},
\end{array}$$
where $\overline{\mathbf{k}}_i \neq \mathbf{k}_i$,
$\overline{\mathbf{k}}'_i \neq \mathbf{k}'_i$.

As explained in Sec.~II, in the present case, the subsystems are
single particle basis states,  and the reduced density matrices
and the entanglement are those of occupation-numbers. For example,
the reduced density matrix of the occupation-number of
$|ci\mathbf{k}_i\rangle$ is
\begin{equation}
\langle n |\rho_{ci\mathbf{k}_i}|n'\rangle = \sum_{n_l\cdots
n_{\infty}} \langle n,n_l\cdots n_{\infty}
|\Psi\rangle\langle\Psi|n',n_l\cdots n_{\infty}\rangle,
\label{rhoci}
\end{equation}
where $l,\cdots,\infty$ represent all the single particle basis
states other than $|ci\mathbf{k}_i\rangle$.

The entanglement between $|ci\mathbf{k}_i\rangle$ and the rest of
the system is thus, as the von Neumann entropy of (\ref{rhoci})
$$S_{ci\mathbf{k}_i} = -
\alpha_{i\mathbf{k}_i}\ln\alpha_{i\mathbf{k}_i}
-(1-\alpha_{i\mathbf{k}_i})\ln(1-\alpha_{i\mathbf{k}_i}).$$
This is obtained by considering $\langle 1
|\rho_{ci\mathbf{k}_i}|1\rangle = \alpha_{i\mathbf{k}_i} =
|g_{Si}|^2\sum_{\mathbf{k}'_i}|\Phi_i(\mathbf{k}_i,\mathbf{k}'_i)|^2$,
$\langle 0 |\rho_{ci\mathbf{k}_i}|0\rangle = 1-
\alpha_{i\mathbf{k}_i}$, and that $\rho_{ci\mathbf{k}_i}$ is
diagonal in the basis $(|0\rangle, |1\rangle)$, basically for the
reason that whenever $|ci\mathbf{k}_i\rangle$ is occupied, there
is always an occupied hole band state.  Similarly, the
entanglement between $|hi\mathbf{k}_i'\rangle$ and the rest of the
system is
$$ S_{hi\mathbf{k}'_i} = -
\alpha_{i\mathbf{k}'_i}\ln\alpha_{i\mathbf{k}'_i}
-(1-\alpha_{i\mathbf{k}'_i})\ln(1-\alpha_{i\mathbf{k}'_i}),$$
where $\alpha_{i\mathbf{k}'_i} =
|g_{Si}|^2\sum_{\mathbf{k}_i}|\Phi_i(\mathbf{k}_i,\mathbf{k}'_i)|^2$.
The entanglement between
$|ci\mathbf{k}_i\rangle|hi\mathbf{k}_i'\rangle$ and the rest of
the system can be calculated to be
$$
\begin{array}{ll}
S_{ci\mathbf{k}_i,hi\mathbf{k}'_i}=&
-|g_{Si}\Phi_i(\mathbf{k}_i,\mathbf{k}'_i)|^2
\ln|g_{Si}\Phi_i(\mathbf{k}_i,\mathbf{k}'_i)|^2 \\
&- \gamma_{i\mathbf{k}_i}\ln\gamma_{i\mathbf{k}_i}-
\gamma_{i\mathbf{k}'_i}\ln\gamma_{i\mathbf{k}'_i}
\\& -(1-\gamma_{i\mathbf{k}_i}-\gamma_{i\mathbf{k}'_i})
\ln(1-\gamma_{i\mathbf{k}_i}-\gamma_{i\mathbf{k}'_i}),
\end{array}$$ where $\gamma_{i\mathbf{k}_i} =
|g_{Si}|^2\sum_{\mathbf{q}'_i\neq
\mathbf{k}_i'}|\Phi_i(\mathbf{k}_i,\mathbf{k}'_i)|^2$,
$\gamma_{i\mathbf{k}'_i} = |g_{Si}|^2\sum_{\mathbf{q}_i\neq
\mathbf{k}_i}|\Phi_i(\mathbf{k}_i,\mathbf{k}'_i)|^2$.

Note that these three  results  are valid no matter whether the
$g_G$ and $g_{Sj}$ ($j\neq i$) are $0$ or not, which only affects
the value of $1-\alpha_{i\mathbf{k}_i}$ and
$1-\alpha_{i\mathbf{k}'_i}$. When $g_{Sj}=0$, no matter whether
$g_G =0$, the single particle basis states with index $j$ become
separated out. Replacing $\Phi_i(\mathbf{k}_i,\mathbf{k}'_i)$ by
$\Psi_i(\mathbf{r},\mathbf{r})$, one  obtains entanglements
concerning  $|ci\mathbf{r}\rangle$ and $|hi\mathbf{r}'\rangle$,
i.e. when the modes are defined by positions rather than wave
vectors.

{\em When and only when $g_G=0$ while both $g_{S1}$ and $g_{S2}$
are nonzero}, the nature of entanglement can be accounted in terms
of two existing distinguishable particles: {\em one} electron and
{\em one} hole. In the present case, the basis states of the
electron and the hole are spinors. There are two degrees of
freedoms, the band index (i.e. angular momentum and effective
mass) and the orbit (position or  wavevector). The effective state
of the two distinguishable particles is $$\begin{array}{lll}
|\Psi\rangle & = & \sum\limits_{\mathbf{k},\mathbf{k}'} [g_{S1}
\Phi_1(\mathbf{k},\mathbf{k}')|c1\rangle|h1\rangle \\
& & + g_{S2}
\Phi_2(\mathbf{k},\mathbf{k}')|c2\rangle|h2\rangle]\otimes
|\mathbf{k}\rangle|\mathbf{k}'\rangle \nonumber \\
& = & \int\int [g_{S1}
\Psi_1(\mathbf{r},\mathbf{r}')|c1\rangle|h1\rangle \\
&& +
g_{S2}\Psi_2(\mathbf{r},\mathbf{r}')|c2\rangle|h2\rangle]\otimes
 |\mathbf{r}\rangle|\mathbf{r}' \rangle d^3 r d^3 r',
\end{array}$$
where  $|ci\rangle$, $\mathbf{k}$ and $\mathbf{r}$ are for the
electron, $|hi\rangle$, $\mathbf{k}'$ and $\mathbf{r}'$ are for
the hole. The reduced density matrix of the hole,
 $$\begin{array}{lll} \rho_h & \equiv&   \sum_{\mathbf{k}}\sum_i \langle ci|\langle
\mathbf{k}|\Psi\rangle\langle\Psi|\mathbf{k}\rangle|ci\rangle \\
& \equiv &  \sum_i \int d^3 r \langle ci|\langle
\mathbf{r}|\Psi\rangle\langle\Psi|\mathbf{r}\rangle|ci\rangle\end{array}$$
is
$$\begin{array}{c}
\sum\limits_{\mathbf{k},\mathbf{k}',\mathbf{k}'',i}|g_{Si}|^2
\Phi_i(\mathbf{k},\mathbf{k}') \Phi_i^*(\mathbf{k},\mathbf{k}'')
|hi\rangle\langle hi|\otimes|\mathbf{k}'\rangle \langle
\mathbf{k}''| \\ = \sum_{i}\int d^3 r\int d^3 r'\int d^3 r''
|g_{Si}|^2 \Psi_i(\mathbf{r},\mathbf{r}')
\Psi_i^*(\mathbf{r},\mathbf{r}'') \\
\times |hi\rangle\langle hi|\otimes |\mathbf{r}'\rangle \langle
\mathbf{r}''|\end{array}$$. The entanglement between the electron
and the hole is quantified to be $S_{h} = -tr \rho_h \ln \rho_h$.

The orbital state, obtained by tracing out the band indices, is
$$\begin{array}{lll} \rho^{orbits} &  =  & \sum_{\mathbf{k},\mathbf{k}'} [
|g_{S1}|^2|\Phi_1(\mathbf{k},\mathbf{k}')|^2+
|g_{S2}|^2|\Phi_2(\mathbf{k},\mathbf{k}')|^2] \\
&& \times |\mathbf{k}\rangle
\langle\mathbf{k}|\otimes|\mathbf{k}'\rangle\langle \mathbf{k}'|
\nonumber \\
 & =&  \int\int [|g_{S1}|^2 |\Psi_1(\mathbf{r},\mathbf{r}')|^2+
|g_{S2}|^2 |\Psi_2(\mathbf{r},\mathbf{r}')|^2]\\
&& \times |\mathbf{r}\rangle \langle \mathbf{r}|\otimes
 |\mathbf{r}' \rangle\langle\mathbf{r}' |  d^3 r d^3 r',
\end{array}$$
which is presumably a bipartite mixed state of continuous
variables, with each part living in an infinite dimensional
Hilbert space, for which  there is not yet an analytical
entanglement measure.

The most interesting and experimentally detectable  entanglement,
which is indeed the one detected in the \cite{chen}, is that
between the band indices, after the Bloch wavevector or position
wavefunction is traced out. The density matrix of the band-index
state is thus $$\begin{array}{lll} \rho^{bands} & =&
|g_{S1}|^2|h1\rangle\langle h1|\otimes|c1\rangle\langle c1| \\
&& + g_{S1}g_{S2}^* x|h1\rangle\langle h2|\otimes|c1\rangle\langle
c2|
\\ & &+ g_{S2}g_{S1}^* x^*|h2\rangle\langle h1|\otimes|c2\rangle\langle
c1|\\ && +|g_{S2}|^2|h2\rangle\langle h2|\otimes|c2\rangle\langle
c2|,
\end{array}$$
where
$x\equiv\sum_{\mathbf{k},\mathbf{k}'}\Phi_1(\mathbf{k},\mathbf{k}')
\Phi_2^*(\mathbf{k},\mathbf{k}') \equiv\int
\Psi_1(\mathbf{r},\mathbf{r}')\Psi_2^*(\mathbf{r},\mathbf{r}')d^3
r d^3 r'$ is the overlap between the wavefunctions of the two
excitonic eigenstates. Presumably, $\rho^{bands}$, like
$\rho^{orbits}$, is also a mixed state.

However, interestingly
$$\Psi_1(\mathbf{r},\mathbf{r}')=\Psi_2(\mathbf{r},\mathbf{r}')
\equiv\Psi(\mathbf{r},\mathbf{r}'),$$ and
$$\Phi_1(\mathbf{k},\mathbf{k}') =
\Phi_2(\mathbf{k},\mathbf{k}') \equiv
\Phi_(\mathbf{k},\mathbf{k}'),$$ because both $\Psi_1$ and
$\Psi_2$, or both $\Phi_1$ and $\Phi_2$,  are the lowest bound
state wavefunctions, {\em which is independent of the effective
masses}, which only affect the energy. Hence $|x|=1$.

Therefore, {\em both $\rho^{orbits}$ and $\rho^{bands}$ become
pure states}. In other words, in the state
$g_{S1}|S1\rangle+g_{S2}|S2\rangle$, {\em band index and orbital
degrees of freedom become separable}. Consequently, the total
entanglement between the electron and the hole is the sum of the
entanglement in the orbital and that in  band-index states.

The orbital state is $\rho^{bands} = |\phi\rangle\langle\phi|$,
with
$$
\begin{array}{ll}
|\phi\rangle  & =  \sum_{\mathbf{k},\mathbf{k}'}
\Phi(\mathbf{k},\mathbf{k}')|\mathbf{k}\rangle|\mathbf{k}'\rangle
\nonumber\\
 & = \int\int
 \Psi(\mathbf{r},\mathbf{r}')|\mathbf{r}\rangle|\mathbf{r}' \rangle
 d\mathbf{r} d\mathbf{r}', \nonumber
 \end{array}
 $$
in which the entanglement is quantified as the von Neumann entropy
of the reduced density matrix  of either the electron or the hole
obtained from the orbital wavefunction.

The band index state is $\rho^{bands}=|\psi\rangle\langle\psi|$,
with
\begin{equation}
|\psi\rangle =g_{S1}|c1\rangle|h1\rangle + g_{S2}
|c2\rangle|h2\rangle. \label{band}
\end{equation}
Hence the  band-index entanglement between the electron and the
hole is
$$S^{bands}=-|g_{S1}|^2\ln|g_{S1}|^2-|g_{S2}|^2\ln|g_{S2}|^2.$$ A
speciality here is that the state is a superposition of two
eigenstates with different angular momenta, which can be probed by
using magnetic field, as well as  different effective masses,
which can be probed by using cyclotron resonance.

The factorization, or disentanglement, of band index and orbital
states as realized in this state, is very interesting for quantum
computing in semiconductors. If the spin is used as qubit,
spin-orbit coupling causes decoherence and error. But if the total
angular momentum is used as qubit, spin-orbit coupling may not
cause decoherence, as exemplified by the study here.

\section{Spatial separation}

We can  spatially separate the electron and hole by engineering
the orbital envelope wavefunctions of the excitonic eigenstate.
The spatial separation is of significant interest in quantum
information and quantum foundations. Note that the band-index, or
angular-momentum, is not coupled to the external barrier or
electric field which are used in  engineer the orbital
wavefunction.

If the orbital degree of freedom is ``entangled'' with the band
index, then engineering orbital wavefunction also influences the
band index state.  Moreover, it causes problem in whether one can
measure band index state $\rho^{bands}$, which is obtained by
tracing out the orbital degree of freedom. This could be a source
of decoherence of the band-index state. Other sources of
decoherence include the phonons,  nuclear spins, etc.

However, as discussed in the preceding section,  in the state
$g_{S1}|S1\rangle+g_{S2}|S2\rangle$, band index and orbital
degrees of freedom are separated. Hence in spatially separating
the electron and hole, the angular momentum, i.e.  the band index
state is not influenced, simply like the case of spatially
separating an Einstein-Podolsky-Rosen-Bohm pair which is
spin-entangled.  Also, it of course does not matter if the state
has a ground-state component, which is simply not affected.
Therefore  spin-orbit coupling does not cause decoherence in the
band-indices or angular momenta in the state $|\Psi\rangle  = g_G
|G\rangle+ g_{S1}|S1\rangle + g_{S2}|S2\rangle$.

In the following, we suggest a few methods of achieving spatial
separation, by exploiting various physical properties of
semiconductor heterostructures~\cite{basu}. One method is to let
the quantum dot or well, in which the electron-hole pair is
generated, tunnel-couple with another one or more dots or wells.
When the size of the dot or well is no smaller than the radius of
the two excitonic eigenstates, the optical generation is not
affected. After generation, tunnelling gives rise to probability
of finding electron and hole in different dots.

It is intriguing to give some detail of the tunnelling of the
entangled state. The total Hamiltonian is
$H = {\cal H}_A+{\cal H}_B+ H_T$,
where ${\cal H}_A$ and ${\cal H}_B$ are electronic Hamiltonians in
the two dots. The tunnelling Hamiltonian is
$$\begin{array}{lll}
H_{T} & =& \sum_i\sum_{\mathbf{k}_{\mu}\mathbf{k}_{\nu}}
(t_{ci\mathbf{k}_{\mu}\mathbf{k}_{\nu}}
{a^A_{i\mathbf{k}_{\mu}}}^{\dagger}a^B_{i\mathbf{k}_{\nu}} +H.c)\\
&&+\sum_i\sum_{\mathbf{k}_{\mu}\mathbf{k}_{\nu}}
(t'_{hi\mathbf{k}_{\mu}\mathbf{k}_{\nu} }
{b^A_{i-\mathbf{k}_{\mu}}}^{\dagger}b^B_{i-\mathbf{k}_{\nu}}+H.c).
\end{array}$$
Because it does not change the band index, tunnelling changes the
overall state through the change of the envelope function of {\em
each} excitonic eigenstate independently, from an excitonic bound
state to a superposition including the component in which the
electron and hole reside in different dots.  Suppose the optically
generated state is given by (\ref{ptt}).  With tunnelling, the
state can still be written in the form of (\ref{ptt}), with only
the orbital wavefunctions of $|S1\rangle$ and $|S2\rangle$
transformed. The band-index state remains unaffected. If during
the tunnelling, the magnetic field which removes the band
degeneracy is present or absent in both dots or wells, then the
two conduction bands see a same barrier, and the two hole bands
also see a same barrier. Consequently, given
$\Psi_1(\mathbf{r},\mathbf{r}')$ and
$\Psi_2(\mathbf{r},\mathbf{r}')$ are equal initially, they remain
equal under tunnelling, though each becomes a delocalized
superposition. When $g_G=0$, this gives rise to spatial separation
of the electron and the hole in the band-index pure state
(\ref{band}). Remember there is only one pair of electron and
hole, which is in a superposition state before measurement or
decoherence.

Furthermore, an electric field can localize electron and hole in
different dots or wells, due to Wannier-Stark effect.
Consequently,
$\Psi_1(\mathbf{r},\mathbf{r}')=\Psi_2(\mathbf{r},\mathbf{r}')$
becomes $\phi_A(\mathbf{r})\phi_B(\mathbf{r}')$. The electric
field may either be exerted after optical generation or be present
even during  the optical generation. The latter option, however,
shifts the resonant energies and, when the field is strong enough,
causes ionization, which then brings  in the bi-excitonic
component. Note that electric field does not couple to the
band-index (angular momentum) degree of freedom, and that the
band-index state is separated from the orbital state, therefore
the electric field does not cause the decoherence of the
band-index state.

It is interesting to study entangled electron-hole state in a
superlattice, i.e. many coupled quantum wells. Without electric
field, each excitonic state is delocalized over a large region. By
using an electric field, localization of electron and hole with
large spatial separation can be achieved, allowing various studies
of entanglement properties and quantum informational process.

Another method is to make the two excitonic eigenstates indirect
in real space, i.e. electron and hole are confined in different
sides of the heterojunction. As is well known, this can be
achieved by the so-called type-II heterojunctions,  in which the
lower conduction band and the higher hole band are on the two
different sides. Such an interface is formed by III-V compounds
with both different group III elements and different group V
elements.

\section{summary}

To summarize, in the framework of quantum field theory, we studied
characterizations and optical generation of entanglement in an
electron-hole system, with the consideration of spin-orbit
coupling,  and have given a theoretical account of an interesting
experimental result~\cite{chen}.  For a many-electron system,
different single particle states are distinguishable subsystems.
The entanglement is between occupation-numbers of different single
particle states, and is generated when the field theoretic
Hamiltonian couples different single particle basis states.

For a semiconductor, coupling with two resonant light modes of
different circular polarizations leads to a superposition of
ground state and the two different single excitonic eigenstates,
each of which is made up of an electron and a hole in the
corresponding conduction and heavy-hole bands. In this state,
there exists complicated occupation-number entanglement, which we
have analyzed in detail.

The Coulomb interaction is not essential in generating the
occupation-number entanglement. Occupation-number entanglement
also exists in each excitonic eigenstate, as well as its
superposition with the ground state.

When the state is a superposition of only the two single-excitonic
states, the entanglement can be accounted as between two
distinguishable particles, each with two degrees of freedom, band
index and the orbital degree of freedom. We find that in this
state, tracing out the orbital degree of freedom leads to a pure
entangled state in band-index, and vice versa. Hence in this case,
the band-index and orbital degrees freedom are separated or
non-entangled, despite the spin-orbit coupling in the Hamiltonian.
This finding is interesting for quantum computing in
semiconductors. It suggests that the problem of spin decoherence
due to spin-orbit coupling may be avoided by using the total
angular momenta to encode quantum information.

We also  briefly propose several methods to spatially separate the
electron and the hole, which makes the band-index entanglement
nonlocal and thus allows further manipulations. Band-index
entanglement means entanglement in both angular momenta and
effective masses. This speciality is a consequence of spin-orbit
coupling, hence is a manifestation of relativistic effect on
quantum entanglement, which is also studied in a different
context~\cite{peres}.

Finally, we mention that our method of characterizing the
entanglement and its generation in electron-hole systems can
equally be applied to the processes proposed in Ref.~\cite{bee}.

\acknowledgements

I am grateful to Peter Littlewood  and Tony Leggett for useful
discussions and suggestions.


\begin{thebibliography}{99}

\bibitem{epr} E. Einstein, B. Podolsky and N. Rosen, Phys. Rev. {\bf 47},
777 (1935);  E. Schr\"{o}dinger, Proc. Camb. Phi. Soc. {\bf 31},
555 (1935);  D. Bohm, {\em Quantum Theory} (Prentice-Hall,
Englewood Cliffs, 1951); J. S. Bell, Physics {\bf 1}, 195 (1964).

\bibitem{book1} For reviews, see, e.g.  D. Bouwmeester,
A. Ekert and A. Zeilinger (ed.), {\em The Physics of Quantum
Information} (Springer,2000).

\bibitem{electron}
J. C. Egues {\it et al.}, in  "Quantum Noise in Mesoscopic
Physics", NATO Science Series, vol. 97 (Kluwer, Netherlands,
2003), and references therein; T. Martin, A. Crepieux, and N.
Chtchelkatchev, in {\em Proceedings of the NATO ARW workshop on
Quantum Noise}, ed. Y. Nazarov and Y. Blanter (Kluwer,
Netherlands, 2002), and references therein. G. B. Lesovik, T.
Martin and G. Blatter, Eur. Phys. J. B {\bf 24}, 287 (2001). A. T.
Costa Jr.  and S. Bose, Phys. Rev. Lett. {\bf 87}, 277901 (2001).
W. D. Oliver, F. Yamaguchi, and Y. Yamamoto, Phys. Rev. Lett. {\bf
88}, 037901 (2002). C. Bena, S. Vishveshwara, L. Balents and M. P.
A. Fisher, Phys. Rev. Lett. {\bf 89}, 037901 (2002).  P.
Samuelsson, E. V. Sukhorukov and M. B\"{u}ttiker, Phys. Rev. Lett.
{\bf 91}, 157002 (2003).

\bibitem{chen} G. Chen {\it et al.}, Science {\bf 289}, 1906 (2000).

\bibitem{hohenester} U. Hohenester, Phys. Rev. B {\bf 66}, 245323 (2002).

\bibitem{bee} C. W. J. Beenakker {\it et al.}, Phys.Rev.Lett. {\bf 91},
147901 (2003).

\bibitem{bayer} M. Bayer {\it et al.}, Science {\bf 291}, 451 (2001).
Y. N. Chen, D. S. Chu and T. Brandes, Phys. Rev. Lett. {\bf 90},
166802 (2003).

\bibitem{sham} P. Chen, C. Piermarocchi and L. J. Sham,
cond-mat/0009307. P. Chen, C. Piermarocchi and L. J. Sham, Phys.
Rev. Lett. 87, 067401 (2001). C. Piermarocchi {\it et al.}, Phys.
Rev. B 65, 075307 (2002). P. Chen {\it et al.}, cond-mat/0301422.

\bibitem{exciton}
A. Imamoglu {\it et al.}, Phys. Rev. Lett. {\bf 83}, 4204 (1999).
F. Troiani, U. Hohenester and E. Molinari, Phys. Rev. B {\bf 62}
R2263 (2000). E. Biolatti {\it et al.} Phys. Rev. Lett. {\bf 85},
5647(2000).

\bibitem{johnson} L. Quiroga and N. F. Johnson, Phys.
Rev. Lett. {\bf 83}, 2270 (1999). J. H. Reina, N. F. Johnson,
Phys. Rev. A {\bf 63}, 012303 (2000). F. J. Rodriguez, L. Quiroga,
N. F. Johnson,  Physica Status Solidi (a) {\bf 178} (1), 403-407
(2000). J. H. Reina, Luis Quiroga, N. F. Johnson, Phys. Rev. A
{\bf 62}, 12305 (2000). P. Zhang {\it et al.}, Phys. Rev. A {\bf
67}, 012312 (2003).

\bibitem{li} X. Li et al., Science {\bf 301}, 809 (2003).

\bibitem{shi1} Y. Shi, quant-ph/0204058, J. Phys. A {\bf 37},
6807 (2004); quant-ph/0205069, Phys. Lett. A {\bf 309}, 254
(2003).

\bibitem{shi3} Y. Shi, quant-ph/0205069,
Phys. Rev. A {\bf 67}, 024301 (2003).

\bibitem{zanardi} P. Zanardi, Phys. Rev. A. {\bf 65}, 042101 (2002).

\bibitem{van} S. J. van Enk, Phys. Rev. A 67, 022303 (2003).

\bibitem{photons} Use of parametric
down-conversion is reviewd by N. Gisin, J. G. Rarity and G. Weihs,
in \cite{book1}; Y. H. Shih, Rep. Prog. Phys. {\bf 66}, 1009
(2003). Use of linear optics is discussed in E. Knill, R. Laflamme
and G. J. Milburn, Nature {\bf 409}, 46 (2001).

\bibitem{bennett} C. H. Bennett, H. J. Bernstein, S. Popescu and
B. Schumacher, Phys. Rev. A {\bf 53}, 2046 (1996).

\bibitem{basu} For example, P. K. Basu, {\em Theory of Optical
Processes in Semiconductors} (Clarendon Press, Oxford, 1997).

\bibitem{peres} A. Peres and D. R. Terno, quant-ph/0212023.
A. Peres, P. F. Scudo and D. R. Terno, Phys. Rev. Lett. {\bf 88},
230402 (2002). P. M. Alsing and G. J. Milburn, quant-ph/0203051.
R. M. Gingrich and C. Adami, Phys. Rev. Lett. {\bf 89}, 270402
(2002).

\end{thebibliography}
\end{document}